\documentclass[paper]{ieice}
\usepackage[dvips]{graphicx}
\usepackage{theorem}
\usepackage{amsmath}
\usepackage{amssymb}
\usepackage{graphicx}
\usepackage{subfigure}
\usepackage{color}
\theorembodyfont{\rmfamily}    
\newtheorem{theorem}{Theorem}
\newtheorem{lemma}[theorem]{Lemma}

\newtheorem{corollary}[theorem]{Corollary}

\newtheorem{proposition}[theorem]{Proposition}
\newtheorem{remark}[theorem]{Remark}
\newcommand{\qed}{\hfill$\square$}

\newcommand{\markov}{\leftrightarrow}

\newcommand{\textchange}[1]{#1}

\field{A}
\SpecialIssue{SITA}
\title{Secret Key Agreement from Correlated Gaussian Sources
by Rate Limited Public Communication}

\titlenote{A part of this paper was presented at the 32nd symposium on information theory and its application.}
\authorlist{
 \authorentry[shun-wata@is.tokushima-u.ac.jp]{Shun Watanabe}{m}{labelA}
 \authorentry[oohama@is.tokushima-u.ac.jp]{Yasutada Oohama}{m}{labelA}
}
\affiliate[labelA]{The authors are with the Department of Information Science and Intelligent Systems, University of Tokushima}

\received{2010}{2}{1}
\revised{2010}{6}{25}

\begin{document}
\maketitle
\begin{summary}
We investigate the secret key agreement
from correlated Gaussian sources in which the legitimate
parties can use the public communication with limited rate.
For the class of protocols with the one-way public communication,
we show a closed form expression of
the optimal trade-off between
the rate of key generation and the rate of the public 
communication. Our results clarify an essential difference
between the key agreement from discrete sources and that
from continuous sources.
\end{summary}

\section{Introduction}

Key agreement is one of the most important 
problems in the cryptography, and it has been
extensively studied in the information theory 
for discrete sources (e.g.~\cite{ahlswede:93,csiszar:00,csiszar:04})
since the problem formulation
by Maurer \cite{maurer:93}.
Recently, the confidential message 
transmission \cite{wyner:75,csiszar:78} in 
the MIMO wireless communication has attracted 
considerable attention  as a
practical problem setting 
(e.g.~\cite{liang:08,liu:09,bustin:09}). 
Although the key agreement in the context of the
wireless communication has also attracted
considerable attention recently \cite{bloch:08},
the key agreement from analog sources has not
been studied sufficiently compared to
the confidential message transmission.
As a fundamental case of the key agreement
from analog sources, we consider the key
agreement from correlated Gaussian
sources in this paper. More specifically,
we consider the problem in which the legitimate 
parties, Alice and Bob, and an eavesdropper, Eve,
have correlated Gaussian sources respectively,
and Alice and Bob share a secret key from their sources
by using the public communication.
Recently, the key agreement from Gaussian sources
has attracted considerable attention in the context
of the quantum key distribution \cite{grosshans:03},
which is also a motivation to investigate
the present problem.

Typically, the first step of the key agreement protocol
from analog sources is the quantization of the sources.
In the literatures (e.g.~see \cite{bloch:08,assche:04,assche:book}), 
\textchange{the scalar quantizer is used}, 
i.e., the observed source is quantized in each time instant.
Using the finer quantization, we can expect the higher key
rate in the protocol, where the key rate is the ratio between the length of
the shared key and the block length of
the sources that are used in the protocol.
However, there is a problem such that the finer quantization
might increase the rate of the public communication in the protocol.
Although the public communication is usually regarded as a cheap resource
in the context of the key agreement problem, 
it is limited by a certain amount in practice.
Therefore, we consider the key agreement protocols with 
the rate limited public communication in this paper.
The purpose of this paper is to clarify the optimal trade-off
between the key rate and the public communication rate
of the key agreement protocol from 
Gaussian sources.
It should be emphasized that we consider the
optimal trade-off among the protocols with
not only the scalar quantizer but also the
vector quantizer.

The key agreement by rate limited public communication
was first studied by Csisz\'ar and 
Narayan for discrete sources \cite{csiszar:00}.
For the class of protocols with the one-way public communication,
they characterized the optimal trade-off between the key rate and
the public communication rate in terms of information
theoretic quantities, i.e., they derived the so-called
single letter characterization.
However, there are two difficulties to extend their 
result to the Gaussian sources.

First, the direct part of the proof in \cite{csiszar:00}
heavily relies on the finiteness of the alphabets
of the sources, and cannot be applied to continuous sources.
We show the direct part by using a method
that is similar to the information spectrum approach \cite{han:book}.

Second, although
the converse part of Csisz\'ar and Narayan's characterization
can be easily extended to continuous sources,
the characterization is not computable because
the characterization involves auxiliary random variables
and the ranges of those random variables are unbounded
for continuous sources.
In this paper, we show that Gaussian auxiliary random
variables are sufficient, and we 
derive a closed form expression of the optimal trade-off.
A key tool in the derivation of the closed form expression is
the entropy power inequality \cite{cover}, which has been
applied to solve the Gaussian multiterminal problems
in the literatures \cite{bergmans:74,cheong:78,ozarow:80,oohama:97}.

There is another work that is related to this paper.
Nitinawarat studied the problem in which Alice and
Bob have correlated Gaussian sources and they share
a secret key from their sources by the public 
communication \cite{nitinawarat:08}.
The problem formulation in this paper can be regarded
as a generalization
of \cite{nitinawarat:08} to the case in which 
Eve also has a Gaussian source.
It should be noted that \cite{nitinawarat:08} considered
the key agreement with the rate limited quantization instead of
the rate limited public communication.

The rest of this paper is organized as follows.
In Section \ref{sec:preliminaries}, we formulate
the problem treated in this paper.
Main results and the outlines of
their proofs are presented
in Section \ref{sec:main}. Conclusions are discussed
in Section \ref{sec:conclusion}, and the details 
of the proofs are presented in Appendices.

\section{Preliminaries}
\label{sec:preliminaries}

Let $X$, $Y$, and $Z$ be zero-mean correlated 
Gaussian sources on the set of real
numbers $\mathbb{R}$
respectively.
Then, let $X^n$, $Y^n$, and $Z^n$ be i.i.d.
copies of $X$, $Y$, and $Z$ respectively.
\textchange{We assume that Alice, Bob, and Eve
know the covariance matrix of $(X,Y,Z)$.}
Throughout the paper, upper case letters indicate
random variables, and the corresponding lower case
letters indicate their realizations.
We use the same notations as \cite{cover}
for the entropy, the mutual information, etc..

Although Alice and Bob can use the public communication
interactively in general, we concentrate on the
class of key agreement protocols in which only Alice 
sends a message to Bob over the public channel\footnote{It should
be noted that the results in this paper is valid for the
class of key agreement protocols in which only
Bob sends a message to Alice.}.
First, Alice computes the message
$C_n$ from $X^n$ and sends the
message to Bob
over the public channel.
Then, she also compute the key $S_n$.
Bob computes the key $S_n^\prime$
from $Y^n$ and $C_n$.

The error probability of the protocol
is defined by 
\begin{eqnarray*}
\varepsilon_n := \Pr\{ S_n \neq S_n^\prime \}.
\end{eqnarray*}
The security of the protocol is measured by
the quantity
\begin{eqnarray}
\label{eq:security-1}
\nu_n := \log|{\cal S}_n| - 
 H(S_n|C_n,Z^n),
\end{eqnarray}
where ${\cal S}_n$ is the range of
the key $S_n$, and $|{\cal S}_n|$ indicates
the cardinality of the set ${\cal S}_n$.

In this paper, we are interested in the trade-off
between the public communication rate $R_p$
and the key rate $R_k$.
The rate pair $(R_p,R_k)$ is defined to 
be achievable if there exists a sequence of
protocols satisfying
\begin{eqnarray}
\label{eq:def-achievability-1}
\lim_{n \to \infty} \varepsilon_n &=& 0, \\
\label{eq:def-achievability-2}
\lim_{n \to \infty} \nu_n &=& 0, \\
\label{eq:def-achievability-3}
\limsup_{n \to \infty} \frac{1}{n} \log|{\cal C}_n| &\le& R_p,\\
\label{eq:def-achievability-4}
\liminf_{n \to \infty} \frac{1}{n} \log|{\cal S}_n| &\ge& R_k,
\end{eqnarray}
where ${\cal C}_n$ is the range of the message $C_n$
transmitted over the public channel.
Then, the achievable rate region is defined as
\begin{eqnarray*}
{\cal R}(X,Y,Z) :=
\{(R_p,R_k) : \mbox{$(R_p,R_k)$ is achievable} \}.
\end{eqnarray*}
The purpose of this paper is to derive
a closed form expression of the
rate region ${\cal R}(X,Y,Z)$.

\section{Main Results}
\label{sec:main}

\subsection{Statement of Results}

In this section, we show a closed form
expression of the rate region
${\cal R}(X,Y,Z)$, which will be proved 
in the next section.
Let 
\begin{eqnarray*}
\Sigma = \left[ \begin{array}{ccc}
\Sigma_x & \Sigma_{xy} & \Sigma_{xz} \\
\Sigma_{yx} & \Sigma_y & \Sigma_{yz} \\
\Sigma_{zx} & \Sigma_{zy} & \Sigma_z
\end{array}\right]
\end{eqnarray*} 
be the covariance matrix of $(X,Y,Z)$.
Throughout the paper, we assume that
$\Sigma$ is positive definite
\textchange{and $\Sigma_{xy} \neq 0$ because 
the key agreement is obviously impossible
if $\Sigma_{xy} = 0$.}
Then, we can
write (see Appendix \ref{derivation-of-gaussian-decompose})
\begin{eqnarray}
X &=& K_{xz} Z + W_1, 
 \label{eq:gauss-decompose-1} \\
Y &=& K_{yx} X + K_{yz} Z + W_2 
 \label{eq:gauss-decompose-2} \\
  &=& (K_{yx} K_{xz} + K_{yz}) Z + K_{yx} W_1 + W_2,
 \label{eq:gauss-decompose-3}
\end{eqnarray}
where $W_1$ and $W_2$ are zero-mean Gaussian
random variables independent of each other, $W_1$
is independent of $Z$, and 
$W_2$ is independent of $(X,Z)$.
The coefficients are given by
\begin{eqnarray*}
K_{xz} = \Sigma_{xz} \Sigma_z^{-1}
\end{eqnarray*}
and 
\begin{eqnarray*}
\left[ \begin{array}{cc}
K_{yz} & K_{yx} 
\end{array} \right]
 = \left[ \begin{array}{cc}
 \Sigma_{yz} & \Sigma_{yx}
 \end{array} \right]
\left[ \begin{array}{cc}
 \Sigma_z & \Sigma_{zx} \\
 \Sigma_{xz} & \Sigma_x 
 \end{array} \right]^{-1}.
\end{eqnarray*}
Furthermore, we also have
\begin{eqnarray}
\label{eq:coeff-1}
\Sigma_{W_1} &=& \Sigma_{x|z}
  = \Sigma_x - K_{xz} \Sigma_{zx}, \\
\label{eq:coeff-2}
\Sigma_{W_2} &=& \Sigma_{y|xz}
  = \Sigma_y - K_{yx} \Sigma_{xy} - K_{yz} \Sigma_{zy}, \\
\label{eq:coeff-3}
\Sigma_{y|z}  &=& \Sigma_y - 
  \Sigma_{yz} \Sigma_z^{-1} \Sigma_{zy},
\end{eqnarray}
where $\Sigma_{W_1}$ and $\Sigma_{W_2}$ are
the variances of $W_1$ and $W_2$ respectively,
\textchange{$\Sigma_{x|z}$ is the conditional variance of $X$ given $Z$,
$\Sigma_{y|xz}$ and $\Sigma_{y|z}$ are the conditional
variances of $Y$ given $(X,Z)$ and $Z$ respectively.}

For $R_p \ge 0$, let 
\begin{eqnarray*}
R_k(R_p) := \sup\{ R_k : (R_k,R_p) \in {\cal R}(X,Y,Z) \}.
\end{eqnarray*}
Before investigating the rate region, we 
present well known upper bound on the function
$R_k(R_p)$, which was shown for
the discrete sources in \cite{ahlswede:93,maurer:93}, and
can be shown in a similar manner for continuous sources.
\begin{proposition}
(\cite{ahlswede:93,maurer:93})
\label{proposition:general-upper}
For any $R_p \ge 0$, we have
\begin{eqnarray}
\label{eq:general-upper-bound}
R_k(R_p) \le I(X;Y|Z) = \frac{1}{2}\log\frac{\Sigma_{y|z}}{\Sigma_{y|xz}}.
\end{eqnarray}
\end{proposition}
\begin{remark}
Although we \textchange{will} concentrate on the
class of key agreement protocols in which only Alice 
sends a message to Bob over the public channel,
the upper bound in Proposition \ref{proposition:general-upper}
is still valid even if we consider the class of protocols in which
Alice and Bob sends messages interactively.
\end{remark}
\begin{remark}
\textchange{
In Eq.~(\ref{eq:general-upper-bound}),
we use the fact that $(X,Y,Z)$ are Gaussian
only to derive the right equality, and the left inequality
holds for any continuous sources.}
\end{remark}

We first consider the case such that
the sources are degraded, i.e., they
satisfy the Markov chain
\begin{eqnarray*}
X \markov Y \markov Z.
\end{eqnarray*}
\begin{theorem}
\label{theorem:degraded}
Suppose that $(X,Y,Z)$ are degraded.
Then, we have
\begin{eqnarray}
\lefteqn{
{\cal R}(X,Y,Z) = 
\left\{ (R_p,R_k): 
\phantom{\frac{\Sigma_{y|z}\left(1- e^{-2R_p} \right)}{\Sigma_{y|xz}}} 
\right. } \nonumber \\
&& \hspace{-2mm}
  \left. R_k \le \frac{1}{2}\log \frac{\Sigma_{y|xz} e^{-2R_p} +
   \Sigma_{y|z}\left(1- e^{-2R_p} \right)}{ \Sigma_{y|xz}}
\right\}.
\label{eq:achievable-region}
\end{eqnarray}
\end{theorem}
As we can find from the above theorem, the
function $R_k(R_p)$
is concave and monotonically increasing, and it 
converges to the upper bound in 
Proposition \ref{proposition:general-upper}
as $R_p$ goes to infinity.
\begin{remark}
\label{remark:essential-difference}
When $(X,Y,Z)$ are discrete sources
and are degraded, it is known \cite{ahlswede:93,csiszar:00}
that
\begin{eqnarray*}
R_k(R_p)
 = I(X;Y|Z).
\end{eqnarray*}
\textchange{for any $R_p \ge H(X|Y)$.
Furthermore for $R_p \ge H(X|Y)$, $R_k(R_p)$ can be achieved by
the combination of the Slepian-Wolf coding \cite{slepian:73}
and the privacy amplification 
(e.g.~see \cite{bennett:95}), and the quantization
of Alice's source is not necessary.}
On the other hand, Theorem \ref{theorem:degraded}
implies 
\begin{eqnarray*}
R_k(R_p)
 < I(X;Y|Z)
\end{eqnarray*}
for any finite $R_p$. This fact suggests an
essential difference between the key agreement
from discrete sources and that from
continuous sources.
\end{remark}

When we consider the protocol with only one-way
public communication, note that
the error probability $\varepsilon_n$ and
the security parameter $\nu_n$ only
depend on the marginal densities $p(x,y)$ and $p(x,z)$
respectively. More precisely, let $(\bar{X},\bar{Y},\bar{Z})$
be random variables such that the marginal densities
of $(X,Y)$ and $(\bar{X},\bar{Y})$, and those
of $(X,Z)$ and $(\bar{X},\bar{Z})$ coincide
respectively. Then we have
\begin{eqnarray*}
{\cal R}(X,Y,Z) = {\cal R}(\bar{X},\bar{Y},\bar{Z}).
\end{eqnarray*}
By using this fact and the following lemma, the proof
of which will be presented in Appendix \ref{proof-of-lemma:degradable},
we can always reduce the general case to the degraded case.
\begin{lemma}
\label{lemma:degradable}
\textchange{If the square of
the correlation coefficient of $(X,Y)$ 
is larger than that of $(X,Z)$, i.e.,}
\begin{eqnarray}
\label{case-1}
\Sigma_{xy}^2 \Sigma_y^{-1} \Sigma_x^{-1}
> \Sigma_{xz}^2 \Sigma_z^{-1} \Sigma_x^{-1},
\end{eqnarray}
then there exist jointly Gaussian sources 
$(\bar{X},\bar{Y},\bar{Z})$ such that
\begin{eqnarray}
\label{eq:degradable-1}
\bar{X} \markov \bar{Y} \markov \bar{Z}
\end{eqnarray}
is satisfied, and that the marginal densities
of $(X,Y)$ and $(\bar{X},\bar{Y})$, and those of
$(X,Z)$ and $(\bar{X},\bar{Z})$ coincide respectively.

On the other hand, 
\textchange{if the square of
the correlation coefficient of $(X,Y)$ is smaller
than or equal to
that of $(X,Z)$, i.e.,}
\begin{eqnarray}
\label{case-2}
\Sigma_{xy}^2 \Sigma_y^{-1} \Sigma_x^{-1}
\le \Sigma_{xz}^2 \Sigma_z^{-1} \Sigma_x^{-1},
\end{eqnarray}
then there exist (not necessarily Gaussian) sources
$(\bar{X},\bar{Y},\bar{Z})$ such that
\begin{eqnarray}
\label{eq:degradable-2}
\bar{X} \markov \bar{Z} \markov \bar{Y}
\end{eqnarray}
is satisfied, and that the marginal densities
of $(X,Y)$ and $(\bar{X},\bar{Y})$, and those of
$(X,Z)$ and $(\bar{X},\bar{Z})$ coincide respectively.
\end{lemma}
When there are jointly Gaussian sources
$(\bar{X},\bar{Y},\bar{Z})$ 
satisfying Eq.~(\ref{eq:degradable-1}),
we can compute the region by using
Theorem \ref{theorem:degraded}.
On the other hand, when there are
$(\bar{X},\bar{Y},\bar{Z})$ 
satisfying Eq.~(\ref{eq:degradable-2}),
Proposition \ref{proposition:general-upper}
implies 
$R_k(R_p) = 0$ for any $R_p \ge 0$.


\subsection{Proof of Theorem \ref{theorem:degraded}}

\subsubsection{Converse Part}

In order to prove the converse part,
we need the following proposition and
corollary.
The proposition was shown for discrete
sources in \cite[Theorem 2.6]{csiszar:00},
and it can be shown almost in the same manner
for continuous sources.
\begin{proposition}
(\cite{csiszar:00}) 
\label{prop:cn-bound}
Suppose that a 
rate pair $(R_p,R_k)$ is included in
${\cal R}(X,Y,Z)$. Then, 
there exist
auxiliary random variables $U$ and $V$ satisfying
\begin{eqnarray}
\label{public-lower}
R_p &\ge& I(U;X|Y), \\
R_k &\le& I(U;Y|V) - I(U;Z|V),
\end{eqnarray}
and the Markov chain
\begin{eqnarray}
\label{markov-condition}
V \markov U \markov X \markov (Y,Z).
\end{eqnarray}
\end{proposition}
For degraded sources, we can simplify 
the above proposition,
which will be shown in \ref{app:proof-of-cn-bound}.
\begin{corollary}
\label{coro:cn-bound}
Suppose that $(X,Y,Z)$ is degraded, i.e., $X \markov Y \markov Z$.
If $(R_p,R_k) \in {\cal R}(X,Y,Z)$, then
there exists an auxiliary random variable $U$ satisfying
\begin{eqnarray}
\label{eq:coro-public-upper}
R_p &\ge& I(U;X|Y), \\
\label{eq:coro-key-upper}
R_k &\le& I(U;Y|Z),
\end{eqnarray}
and the Markov chain
\begin{eqnarray}
\label{eq:markov-2}
U \markov X \markov Y \markov Z.
\end{eqnarray}
\end{corollary}

\noindent{\em Proof of Converse Part)}

Part of ideas of the following proof are borrowed 
from \cite{tian:09} with proper modifications.
In the following, $h(\cdot)$ and
$h(\cdot|\cdot)$ designate the
differential entropy and the conditional
differential entropy respectively \cite{cover}.
 
We will show
\begin{eqnarray}
\label{eq:intermidiate}
R_p \ge \frac{1}{2} \log
 \frac{\Sigma_{y|z} - \Sigma_{y|xz}}{
  \Sigma_{y|z} e^{-2R_k} - \Sigma_{y|xz}}
  - R_k.
\end{eqnarray}
Then, by solving the inequality with respect
to $R_k$, we have the converse part
of Theorem \ref{theorem:degraded}.

For any auxiliary random variable $U$
satisfying Eq.~(\ref{eq:markov-2}),
by a straightforward calculation,
we have
\begin{eqnarray*}
\lefteqn{ h(Y|Z) - I(U;Y|Z) } \\
 &=& h(Y|U,Z) \\
 &=& h(K_{yx} X + K_{yz} Z + W_2 |U, Z) \\
 &=& h(K_{yx} X + W_2 | U, Z).
\end{eqnarray*}
Then, by using the conditional version of the 
entropy power inequality (EPI) \cite{bergmans:74},
we have
\begin{eqnarray*}
\lefteqn{
 \exp\left[ 2 h(K_{yx} X + W_2|U,Z) \right] } \\
&\ge& \exp\left[ 2 h(K_{yx} X|U, Z) \right] 
  + \exp\left[ 2 h(W_2) \right] \\
&=& K_{yx}^2 \exp\left[ 2 h(X|U,Z) \right]
  + \exp\left[ 2 h(W_2) \right] \\
&=& K_{yx}^2 \exp\left[ - 2 I(U;X|Z) + 2 h(X|Z) \right] \\
&&  + \exp\left[ 2 h(W_2) \right] \\
&=& K_{yx}^2 \exp\left[ - 2 I(U;X|Z) + 2 h(W_1) \right] \\
&&  + \exp\left[ 2 h(W_2) \right].
\end{eqnarray*}
Thus, we have
\begin{eqnarray}
\lefteqn{ 
 I(U;X|Z) - I(U;Y|Z) } \nonumber \\
&\ge& 
  \frac{1}{2} \log \left[ K_{yx}^2 \exp\{ 2 h(W_1)\} \right]  \nonumber \\
&& - \frac{1}{2} \log\left[ \exp\{2 h(Y|Z) - 2 I(U;Y|Z)\} \right. \nonumber \\ 
 &&    \left. - \exp\{ 2 h(W_2) \} \right]  - I(U;Y|Z).
\label{eq:proof-trade-off-1}
\end{eqnarray}
From Eqs.~(\ref{eq:coeff-2}) and (\ref{eq:gauss-decompose-3}),
we can find that the variances of
$W_2$ and $K_{yx} W_1$ are 
$\Sigma_{y|xz}$ and $K_{yx}^2 \Sigma_{W_1} = \Sigma_{y|z} - \Sigma_{y|xz}$
respectively.
Thus, we can rewrite the right hand side of Eq.~(\ref{eq:proof-trade-off-1})
as 
\begin{eqnarray*}
\frac{1}{2} \log \frac{ \Sigma_{y|z} - \Sigma_{y|xz}}{
   \Sigma_{y|z} e^{ -2 I(U;Y|Z) } - \Sigma_{y|xz} }
  - I(U;Y|Z).
\end{eqnarray*}
Since the function
\begin{eqnarray*}
\frac{1}{2} \log \frac{ \Sigma_{y|z} - \Sigma_{y|xz}}{
   \Sigma_{y|z} e^{ -2 a } - \Sigma_{y|xz} }
  - a
\end{eqnarray*}
is monotonically increasing  
for $0 \le a \le I(X;Y|Z)$ and
\begin{eqnarray*}
I(U;X|Z) - I(U;Y|Z) = I(U;X|Y)
\end{eqnarray*}
for $(U,X,Y,Z)$ satisfying Eq.~(\ref{eq:markov-2}),
Corollary \ref{coro:cn-bound} implies Eq.~(\ref{eq:intermidiate}).
\qed

\subsubsection{Direct Part}

In order to prove the direct part, we 
need the following proposition, which can be 
regarded as a generalization 
of \cite[Theorem 2.6]{csiszar:00}
to continuous sources\footnote{Although
\cite[Theorem 2.6]{csiszar:00} involves two
auxiliary random variables, we only show 
the version with only one auxiliary random variable
because one of the auxiliary random variables
in \cite[Theorem 2.6]{csiszar:00} is not needed 
to show Theorem \ref{theorem:degraded}.}. 
We show a proof
in \ref{app:proof-of-direct}
because the proof of
\cite[Theorem 2.6]{csiszar:00} heavily relies on
the finiteness of the alphabets and its
generalization to continuous sources seems
non-trivial.
It should be noted that the following proposition
holds for non-degraded case.
\begin{proposition}
\label{proposition:direct-part}
For an auxiliary random variable
$U$ satisfying the Markov chain
\begin{eqnarray}
\label{eq:direct-part:markov}
U \markov X \markov (Y,Z),
\end{eqnarray}
let $(R_p,R_k)$ be a rate pair such that
\begin{eqnarray*}
R_p &\ge& I(U;X) - I(U;Y), \\
R_k &\le& I(U;Y) - I(U;Z).
\end{eqnarray*}
Then, we have 
$(R_p,R_k) \in {\cal R}(X,Y,Z)$.
\end{proposition}
\textchange{
Note that 
\begin{eqnarray*}
I(U;X|Y) &=& I(U;X) - I(U;Y), \\
I(U;Y|Z) &=& I(U;Y) - I(U;Z)
\end{eqnarray*}
for degraded sources.}

\noindent{\em Proof of Direct Part)}

\textchange{
Let $W$ be a zero mean Gaussian random variable,
and $U = X + W$. Since the (conditional version of) entropy power inequality
holds with equality for Gaussian random variables \cite{bergmans:74,cover}, 
Eq.~(\ref{eq:proof-trade-off-1}) in the converse part holds with
equality, i.e., we have}

\textchange{
\begin{eqnarray*}
\lefteqn{
I(U;X|Y)
} \\
&=& I(U;X|Z) - I(U;Y|Z) \\
&=& \frac{1}{2} \log \frac{ \Sigma_{y|z} - \Sigma_{y|xz}}{
   \Sigma_{y|z} e^{ -2 I(U;Y|Z) } - \Sigma_{y|xz} }
  - I(U;Y|Z).
\end{eqnarray*}
Thus, by setting $I(U;Y|Z) = R_k$ and $I(U;X|Y) = R_p$,
by solving the equality with respect to $R_k$,
and by adjusting the variance of $W$, we find that
any $(R_p,R_k)$ satisfying the equality
\begin{eqnarray*}
R_k = \frac{1}{2} \log \frac{\Sigma_{y|xz} e^{-2 R_p} + \Sigma_{y|z} (1-e^{-2 R_p}) }{\Sigma_{y|xz}}
\end{eqnarray*}
is achievable. \qed
}
\section{Conclusions and Discussions}
\label{sec:conclusion}

We investigated the secret key
agreement from Gaussian sources by
rate limited public communication.
For the class of protocols with
the one-way public communication,
we derived a closed form expression 
of the optimal trade-off between the key
rate and the public communication rate.
\textchange{The optimal trade-off for
the class of protocols with the two-way
public communication remains unsolved and
investigating it is a future research agenda.}

Our result suggested an essential difference
between the key agreement from discrete sources
and that from continuous sources (Remark \ref{remark:essential-difference}).
\textchange{For discrete sources, if the public communication
rate is larger than $H(X|Y)$, the upper bound 
can be achieved without quantization.
On the other hand for Gaussian sources,
the upper bound cannot be achieved 
for any finite public communication rate.}

The problem formulation treated in this paper 
can be regarded as Gaussian version of the
source type model \cite{ahlswede:93}.
We can also consider Gaussian version of
the channel type model. 
In such a model, Alice can 
send a signal, with power constraint, to
Bob and Eve over Gaussian channels. In 
addition to the Gaussian channel, Alice and
Bob can use the public communication with
limited rate. For the class of protocols
with the forward public communication\footnote{The forward (backward)
public communication means that only Alice (Bob) sends
a public message to Bob (Alice). It should be noted that
the forward public communication and the backward
public communication make significant difference 
for the channel type model.},
by a slight modification
of the proof of \cite[Theorem 2]{ahlswede:93},
we can show that the supremum of achievable
key rates coincides with the secrecy capacity
of the Gaussian wiretap channel \cite{cheong:78}
no matter what the limitation of the public communication
rate.


\appendix
\section{Proof of Proposition \ref{proposition:direct-part}}
\label{app:proof-of-direct}

For arbitrarily fixed auxiliary random
variable satisfying Eq.~(\ref{eq:direct-part:markov}),
we show that the rate pair 
\begin{eqnarray*}
R_p &=& I(U;X) - I(U;Y) + 4 \gamma , \\
R_k &=& I(U;Y) - I(U;Z) - 6 \gamma
\end{eqnarray*}
is achievable for
any $\gamma > 0$.
Instead of showing the achievability for the
security criterion defined by Eq.~(\ref{eq:security-1}),
we show the achievability for the security
criterion defined by
\begin{eqnarray*}
\lefteqn{ \mu_n := \int p(z^n) } \\
&& \| P_{S_n C_n|Z^n}(\cdot,\cdot|z^n) - 
 P_{\bar{S}_n}(\cdot) P_{C_n|Z^n}(\cdot|z^n) \| dz^n,
\end{eqnarray*}
where $P_{\bar{S}_n}$ is the uniform distribution
on the key alphabet ${\cal S}_n$, and
$\| \cdot \|$ is the variational distance \cite{cover}.
More precisely, we show that there exists a sequence
of protocols satisfying 
Eqs.~(\ref{eq:def-achievability-1}),
(\ref{eq:def-achievability-3}) and 
(\ref{eq:def-achievability-4}) and
$\mu_n$ converges to $0$ exponentially.
Then, by using \cite[Lemma 3]{naito:08},
we can also show that $\nu_n$ also 
converges to $0$.

Our protocol roughly consists of three
steps: the quantization, the bin coding \cite{cover},
and the privacy amplification.
First, Alice quantizes her source by
a function $g_n: \mathbb{R}^n \to {\cal Q}_n \subset \mathbb{R}^n$.
\textchange{We use the auxiliary random variable
$U$ for quantization almost in a similar manner as
the Wyner-Ziv problem \cite{wyner:76}.
After the quantization, 
she sends the bin index $C_n = \phi_n(g_n(X^n))$
by a function $\phi_n:{\cal Q}_n \to {\cal C}_n$.
Bob decodes the index and his source by a
function $\psi_n:{\cal C}_n \times \mathbb{R}^n \to {\cal Q}_n$.
Note that the public communication rate $R_p$ must be
large enough so that Bob can recover the quantized source
by using his source $Y^n$ as side-information at the 
decoder.
}
Finally, they obtain keys 
$S_n = f_n(g_n(X^n))$ and 
$S_n^\prime = f_n(\psi_n(\phi_n(C_n),Y^n))$
by a function $f_n:{\cal Q}_n \to {\cal S}_n$
respectively.
Existence of functions
$\{(g_n,\phi_n,\psi_n,f_n)\}_{n=1}^{\infty}$
satisfying Eqs.~(\ref{eq:def-achievability-1}),
(\ref{eq:def-achievability-3}) and 
(\ref{eq:def-achievability-4}) and
$\mu_n \to 0$ are guaranteed by the following
lemmas.
Lemma \ref{lemma:markov} is the so-called
Markov lemma, the proof of which will be
omitted (e.g.~see \cite{iwata:02}).
In order to upper bound the error probability
$\varepsilon_n$, we need
Lemma \ref{lemma:bin-coding}, which
appears in the course of deriving
the general formula of the 
Wyner-Ziv problem \cite{iwata:02}.
We also omit the proof. In order to
upper bound the security parameter $\mu_n$, we need 
Lemma \ref{lemma:privacy-amplification}, which
is an extension of
the privacy amplification lemma shown in
\cite[Lemma 4]{naito:08}.
Since we apply the privacy amplification
to the (vector) quantized source in our protocol,
we need Lemma \ref{lemma:privacy-amplification}.
The proof of Lemma \ref{lemma:privacy-amplification}
is the most difficult part of the proof of
Proposition \ref{proposition:direct-part}, 
and it will be proved in the next section.

For a fixed auxiliary random variable $U$
satisfying Eq.~(\ref{eq:direct-part:markov})
and $t,\alpha,\beta \in \mathbb{R}$,
let
\begin{eqnarray*}
{\cal T}_n &:=&
 \left\{ (u^n,x^n): \frac{1}{n} \log 
 \frac{p(u^n|x^n)}{p(u^n)} \le t \right\}, \\
{\cal A}_n &:=&
 \left\{ (u^n,y^n): \frac{1}{n} \log
  \frac{p(y^n|u^n)}{p(y^n)} \ge \alpha \right\}, \\
{\cal B}_n &:=&
 \left\{ (u^n,x^n,z^n): \frac{1}{n} \log
  \frac{p(x^n|u^n,z^n)}{p(x^n|z^n)}
  \ge \beta \right\}.
\end{eqnarray*}
\begin{lemma}
\label{lemma:markov}
For any $t,\alpha,\beta \in \mathbb{R}$, 
there exists a function $g_n:\mathbb{R}^n \to {\cal Q}_n$
such that 
\begin{eqnarray*}
\lefteqn{ \hspace{-3mm}
 \Pr\{ (g_n(X^n),Y^n) \notin {\cal A}_n \mbox{ or }
 (g_n(X^n),X^n,Z^n) \notin {\cal B}_n \} } \\
 && \hspace{-4mm} \le 2  \sqrt{\delta_n} + 
 \Pr\{(U^n,X^n) \notin {\cal T}_n \}
 + \exp\{- |{\cal Q}_n| e^{-tn} \},
\end{eqnarray*}
where 
\begin{eqnarray*}
\delta_n := \Pr\{
 (U^n,Y^n) \notin {\cal A}_n \mbox{ or }
  (U^n,X^n,Z^n) \notin {\cal B}_n \}.
\end{eqnarray*}
\end{lemma}
\begin{lemma}
\label{lemma:bin-coding}
For any function 
$g_n:\mathbb{R}^n \to {\cal Q}_n$ and
$\alpha \in \mathbb{R}$, there 
exist functions 
$\phi_n: {\cal Q}_n \to {\cal C}_n$
and $\psi_n: {\cal C}_n \times \mathbb{R}^n \to {\cal Q}_n$
such that
\begin{eqnarray*}
\lefteqn{
 \Pr\{ g_n(X^n) \neq \psi_n(\phi_n(g_n(X^n)), Y^n) \} } \\
&\le&
 \frac{|{\cal Q}_n| }{|{\cal C}_n|} e^{-\alpha n}
 + \Pr\{ (g_n(X^n), Y^n) \notin {\cal A}_n \}.
\end{eqnarray*}
\end{lemma}
\begin{lemma}
\label{lemma:privacy-amplification}
For any functions
$g_n:\mathbb{R}^n \to {\cal Q}_n$,
$\phi_n:{\cal Q}_n \to {\cal C}_n$,
 and
$\beta \in \mathbb{R}$, there 
exists a function $f_n:{\cal Q}_n \to {\cal S}_n$
such that
\begin{eqnarray*}
\hspace{-5mm} \mu_n \le
 \sqrt{|{\cal S}_n| |{\cal C}_n| e^{-\beta n} }
 + 2 \Pr\{ (g_n(X^n),X^n,Z^n) \notin {\cal B}_n \}.
\end{eqnarray*}
\end{lemma}
Note that only the cardinality of 
${\cal C}_n$ appears in the upper bound on $\mu_n$
no matter the structure of a specific function $\phi_n$.
However the functions $f_n$ realizing the upper bound
depend on the structure of $\phi_n$.

From Lemmas \ref{lemma:markov},
\ref{lemma:bin-coding}, and \ref{lemma:privacy-amplification},
by setting
\begin{eqnarray*}
|{\cal Q}_n| &=& \exp\{ n(I(U;X) + 2 \gamma) \}, \\
|{\cal C}_n| &=& \exp\{ n(I(U;X) - I(U;Y) + 4 \gamma) \}, \\
|{\cal S}_n| &=& \exp\{ n(I(U;Y) - I(U;Z) - 6 \gamma) \}, \\
t &=& I(U;X) + \gamma, \\
\alpha &=& I(U;Y) - \gamma, \\
\beta &=& I(U;X|Z) - \gamma,
\end{eqnarray*}
and by noting that $I(U;X|Z) = I(U;X) - I(U;Z)$,
we obtain a sequence of protocols satisfying
Eqs.~(\ref{eq:def-achievability-1}),
(\ref{eq:def-achievability-3}) and 
(\ref{eq:def-achievability-4}) and
$\mu_n$ exponentially\footnote{We use the
Chernoff bound (e.g.~see \cite{cover}) instead
of the Chebyshev inequality to upper bound
$\delta_n$ and $\Pr\{(U^n,X^n) \notin {\cal T}_n \}$.} converges
to $0$.
By using \cite[Lemma 3]{naito:08}, we can show 
that Eq.~(\ref{eq:def-achievability-2}) is also satisfied.

Finally, by taking a sequence $\{\gamma_i \}$
such that $\gamma_1 > \gamma_2 > \cdots > 0$ and
$\gamma_i \to 0~(i \to \infty)$ instead of
$\gamma >0$, and
by using the diagonalization
argument \cite{han:book}, we have
Proposition \ref{proposition:direct-part}.
\qed

\subsection{Proof of Lemma \ref{lemma:privacy-amplification}}

In the following, we use the notation
\begin{eqnarray*}
f_n^{-1}(s) := \{ u^n \in {\cal Q}_n :
 f_n(u^n) = s \}.
\end{eqnarray*}
The sets $\phi_n^{-1}(c)$ for $c \in {\cal C}_n$
and $g_n^{-1}(u^n)$ for
$u^n \in {\cal Q}_n$ are defined in similar manners.
Furthermore, for a set $A \subset {\cal Q}_n$,
we denote $g_n^{-1}(A) := \cup_{u^n \in A} g_n^{-1}(u^n)$.
For a set $B \subset \mathbb{R}^n$, we define
\begin{eqnarray*}
P_{X^n|Z^n}(B|z^n) := \Pr\{ X^n \in B | Z^n = z^n\}.
\end{eqnarray*}
In this section,
it should be also noted that
summations are taken over the range of
the indices unless otherwise specified.

For fixed $z^n \in \mathbb{R}^n$, let
\begin{eqnarray*}
{\cal B}_{z^n} := \{ x^n : (g_n(x^n),x^n, z^n) \in {\cal B}_n \},
\end{eqnarray*}
and ${\cal B}_{z^n}^c$ be the complement
of ${\cal B}_{z^n}$ in $\mathbb{R}^n$.
Then, by a straightforward calculation, we have
\begin{eqnarray}
\lefteqn{
 \| P_{S_n C_n|Z^n}(\cdot,\cdot|z^n) - P_{\bar{S}_n}(\cdot)
  P_{C_n|Z^n}(\cdot|z^n) \| } \nonumber \\
&=& \sum_{s,c} |
  P_{S_n C_n|Z^n}(s,c|z^n) - P_{\bar{S}_n}(s)
  P_{C_n|Z^n}(c|z^n) | \nonumber \\
&=& \sum_{s,c} |
  P_{X^n|Z^n}(g_n^{-1}(f_n^{-1}(s) \cap \phi_n^{-1}(c)) |z^n) \nonumber \\
&& - P_{\bar{S}_n}(s) P_{X^n|Z^n}(g_n^{-1}(\phi_n^{-1}(c))|z^n) | \nonumber \\
&=& \sum_{s,c} |P_{X^n|Z^n}(g_n^{-1}(f_n^{-1}(s) \cap \phi_n^{-1}(c))  \cap {\cal B}_{z^n}|z^n) \nonumber \\
&& - P_{\bar{S}_n}(s) P_{X^n|Z^n}(g_n^{-1}(\phi_n^{-1}(c)) \cap {\cal B}_{z^n}|z^n)  \nonumber \\
&& + P_{X^n|Z^n}(g_n^{-1}(f_n^{-1}(s) \cap \phi_n^{-1}(c))  \cap {\cal B}_{z^n}^c|z^n) \nonumber \\
&& - P_{\bar{S}_n}(s) P_{X^n|Z^n}(g_n^{-1}(\phi_n^{-1}(c)) \cap {\cal B}_{z^n}^c|z^n) | \nonumber \\
&\le& \sum_{s,c} |P_{X^n|Z^n}(g_n^{-1}(f_n^{-1}(s) \cap \phi_n^{-1}(c))  \cap {\cal B}_{z^n}|z^n) \nonumber \\
&& - P_{\bar{S}_n}(s) P_{X^n|Z^n}(g_n^{-1}(\phi_n^{-1}(c)) \cap {\cal B}_{z^n}|z^n) | \nonumber \\
&+&  \sum_{s,c} |P_{X^n|Z^n}(g_n^{-1}(f_n^{-1}(s) \cap \phi_n^{-1}(c))  \cap {\cal B}_{z^n}^c|z^n) \nonumber \\
&& - P_{\bar{S}_n}(s) P_{X^n|Z^n}(g_n^{-1}(\phi_n^{-1}(c)) \cap {\cal B}_{z^n}^c|z^n) | \nonumber \\
&=& \sum_{s,c} |P_{X^n|Z^n}(g_n^{-1}(f_n^{-1}(s) \cap \phi_n^{-1}(c))  \cap {\cal B}_{z^n}|z^n) \nonumber \\
&& - P_{\bar{S}_n}(s) P_{X^n|Z^n}(g_n^{-1}(\phi_n^{-1}(c)) \cap {\cal B}_{z^n}|z^n) | \nonumber \\
&& + 2 P_{X^n|Z^n}({\cal B}_{z^n}^c|z^n),
\label{eq:proof-1}
\end{eqnarray}
where we used the triangle inequality.
By using the Cauchy-Schwarz inequality, the first
term of Eq.~(\ref{eq:proof-1}) is upper 
bounded by
\begin{eqnarray}
\lefteqn{ \hspace{-7mm} \left[ |{\cal S}_n| |{\cal C}_n| \sum_{s,c} |
 P_{X^n|Z^n}(g_n^{-1}(f_n^{-1}(s) \cap \phi_n^{-1}(c))  \cap {\cal B}_{z^n}|z^n) \right. } \nonumber \\
&& \hspace{-14mm} \left. \phantom{\sum_{s}} -
 P_{\bar{S}_n}(s) P_{X^n|Z^n}(g_n^{-1}(\phi_n^{-1}(c)) \cap {\cal B}_{z^n}|z^n) |^2 \right]^{1/2}.
\label{eq:proof-2}
\end{eqnarray}
Furthermore, we can rewrite the inside of the square
root of Eq.~(\ref{eq:proof-2}) as
\begin{eqnarray}
\lefteqn{ \hspace{-6mm}
 \sum_{s,c} |
 P_{X^n|Z^n}(g_n^{-1}(f_n^{-1}(s) \cap \phi_n^{-1}(c))  \cap {\cal B}_{z^n}|z^n)
} \nonumber \\
&& \hspace{-6mm} - P_{\bar{S}_n}(s) P_{X^n|Z^n}(g_n^{-1}(\phi_n^{-1}(c)) \cap {\cal B}_{z^n}|z^n) |^2 \nonumber \\
&&  \hspace{-8mm} = \sum_{s,c}
 P_{X^n|Z^n}(g_n^{-1}(f_n^{-1}(s) \cap \phi_n^{-1}(c))  \cap {\cal B}_{z^n}|z^n)^2 \nonumber \\
&& \hspace{-6mm} - \sum_c
 \frac{1}{|{\cal S}_n|}
 P_{\bar{S}_n}(s) P_{X^n|Z^n}(g_n^{-1}(\phi_n^{-1}(c)) \cap {\cal B}_{z^n}|z^n)^2, \nonumber \\
\label{eq:proof-3}
\end{eqnarray}
where we used the facts $P_{\bar{S}_n}(s) = \frac{1}{|{\cal S}_n|}$
and
\begin{eqnarray*}
\lefteqn{
\sum_{s} P_{X^n|Z^n}(g_n^{-1}(f_n^{-1}(s) \cap \phi_n^{-1}(c)) \cap
  {\cal B}_{z^n} | z^n) } \\
&=& P_{X^n|Z^n}(g_n^{-1}(\phi_n^{-1}(c)) \cap {\cal B}_{z^n}|z^n).
\end{eqnarray*}
We can rewrite the first term of 
Eq.~(\ref{eq:proof-3}) as
\begin{eqnarray}
\lefteqn{
\hspace{-6mm} \sum_{s,c}
 P_{X^n|Z^n}(g_n^{-1}(f_n^{-1}(s) \cap \phi_n^{-1}(c))  \cap {\cal B}_{z^n}|z^n)^2
} \nonumber \\
&& \hspace{-9mm} 
= \sum_{s,c} \sum_{u^n,\hat{u}^n \in f_n^{-1}(s) \cap \phi_n^{-1}(c)} \nonumber \\
&& \hspace{-8mm} P_{X^n|Z^n}(g_n^{-1}(u^n) \cap {\cal B}_{z^n}|z^n)
P_{X^n|Z^n}(g_n^{-1}(\hat{u}^n) \cap {\cal B}_{z^n}|z^n) \nonumber \\
&& \hspace{-9mm} 
= \sum_{s,c} \sum_{u^n,\hat{u}^n \in \phi_n^{-1}(c)} 
  \delta_{f_n(u^n),f_n(\hat{u}^n)} \nonumber \\
&& \hspace{-8mm} P_{X^n|Z^n}(g_n^{-1}(u^n) \cap {\cal B}_{z^n}|z^n)
P_{X^n|Z^n}(g_n^{-1}(\hat{u}^n) \cap {\cal B}_{z^n}|z^n), \nonumber \\
\label{eq:proof-4}
\end{eqnarray}
where $\delta_{a,b} = 1$ if $a=b$ and 
$\delta_{a,b} = 0$ otherwise.
In a similar manner,
we can rewrite the second term of 
Eq.~(\ref{eq:proof-3}) as
\begin{eqnarray}
\lefteqn{ \hspace{-7mm}
\sum_c
 \frac{1}{|{\cal S}_n|}
 P_{\bar{S}_n}(s) P_{X^n|Z^n}(g_n^{-1}(\phi_n^{-1}(c)) \cap {\cal B}_{z^n}^c|z^n)^2
} \nonumber \\
&& \hspace{-9mm} = \sum_{c} \frac{1}{|{\cal S}_n|}
 \sum_{u^n,\hat{u}^n \in \phi_n^{-1}(v)} \nonumber \\
&& \hspace{-9mm} P_{X^n|Z^n}(g_n^{-1}(u^n) \cap {\cal B}_{z^n}|z^n)
P_{X^n|Z^n}(g_n^{-1}(\hat{u}^n) \cap {\cal B}_{z^n}|z^n). \nonumber \\
\label{eq:proof-5}
\end{eqnarray}

Let ${\cal F}_n$ be a universal
hash family of functions
from ${\cal Q}_n$ to ${\cal S}_n$ \cite{carter:79}, i.e.,
\begin{eqnarray*}
P_{F_n}(\{ f_n \in {\cal F}_n: f_n(u^n) = f_n(\hat{u}^n) \})
 \le \frac{1}{|{\cal S}_n|}
\end{eqnarray*}
for any distinct $u^n$ and $\hat{u}^n$,
where $P_{F_n}$ is the uniform distribution on ${\cal F}_n$.
Combining Eqs.~(\ref{eq:proof-3})--(\ref{eq:proof-5}),
we can evaluate the inside of the square root of
Eq.~(\ref{eq:proof-2}) averaged over the random 
choice of $f_n$ as follows:
\begin{eqnarray}
\lefteqn{ \hspace{-7mm}
\mathbb{E}_{f_n} \left[
\sum_{s,c} P_{X^n|Z^n}(g_n^{-1}(f_n^{-1}(s) \cap \phi_n^{-1}(c))
  \cap {\cal B}_{z^n}| z^n)^2 
 \right. } \nonumber \\
&& \hspace{-9mm}
\left. - \sum_c \frac{1}{|{\cal S}_n|}
  P_{X^n|Z^n}(g_n^{-1}(\phi_n^{-1}(c)) \cap {\cal B}_{z^n}|z^n)^2
 \right] \nonumber \\
&& \hspace{-9mm} =
 \sum_c \sum_{u^n, \hat{u}^n \in \phi_n^{-1}(c)}
  \mathbb{E}\left[ \delta_{f_n(u^n) ,f_n(\hat{u}^n)}
  - \frac{1}{|{\cal S}_n|} \right] \nonumber \\
&& \hspace{-9mm}
P_{X^n|Z^n}(g_n^{-1}(u^n) \cap {\cal B}_{z^n}|z^n)
  P_{X^n|Z^n}(g_n^{-1}(\hat{u}^n) \cap {\cal B}_{z^n}|z^n) \nonumber \\
&& \hspace{-9mm} \le
 \sum_c \sum_{u^n \in \phi_n^{-1}(c)} \nonumber \\
  && \hspace{-9mm}
P_{X^n|Z^n}(g_n^{-1}(u^n) \cap {\cal B}_{z^n}|z^n)
  P_{X^n|Z^n}(g_n^{-1}(u^n) \cap {\cal B}_{z^n}|z^n) \nonumber \\
&& \hspace{-9mm} =
 \sum_{u^n \in {\cal Q}_n} \nonumber \\
 && \hspace{-9mm}
P_{X^n|Z^n}(g_n^{-1}(u^n) \cap {\cal B}_{z^n}|z^n)
  P_{X^n|Z^n}(g_n^{-1}(u^n) \cap {\cal B}_{z^n}|z^n) \nonumber \\
\label{eq:proof-6}
\end{eqnarray}
Since
\begin{eqnarray*}
p(x^n|z^n) \le
 p(x^n|u^n,z^n)  e^{-\beta n}
\end{eqnarray*}
for $(u^n,x^n,z^n) \in {\cal B}_n$, 
Eq.~(\ref{eq:proof-6}) is upper bounded by
\begin{eqnarray}
\lefteqn{
\sum_{u^n \in {\cal Q}_n}
 P_{X^n|Z^n}(g_n^{-1}(u^n) \cap {\cal B}_{z^n}|z^n) 
} \nonumber \\
&& P_{X^n|U^n Z^n}(g_n^{-1}(u^n) \cap {\cal B}_{z^n}|u^n,z^n)
 e^{-\beta n} \nonumber \\
&\le&  
 \sum_{u^n \in {\cal Q}_n}
 P_{X^n|Z^n}(g_n^{-1}(u^n) \cap {\cal B}_{z^n}|z^n)
 e^{-\beta n} \nonumber \\
&\le&  e^{-\beta n}.
\label{eq:proof-7}
\end{eqnarray}
\textchange{
Since the square root function $\sqrt{\cdot}$
is concave, by combining 
Eqs.~(\ref{eq:proof-3})--(\ref{eq:proof-7}),
Eq.~(\ref{eq:proof-2}) averaged over $f_n$ is upper 
bounded by $\sqrt{|{\cal S}_n| |{\cal C}_n| e^{- \beta n}}$.
By substituting this upper bound into Eq.~(\ref{eq:proof-1}),
by taking the average over $z^n \in \mathbb{R}^n$, and by 
using the concavity of $\sqrt{\cdot}$, we have
\begin{eqnarray*}
\lefteqn{
\mathbb{E}_{f_n}[\mu_n] \le
  \sqrt{|{\cal S}_n| |{\cal C}_n| e^{-\beta n} } } \\
&&  + 2 \Pr\{ (g_n(X^n),X^n,Z^n) \notin {\cal B}_n \}.
\end{eqnarray*}
Thus, there exists at least one $f_n \in {\cal F}_n$
satisfying the statement of the lemma. \qed
}

\section{Proof of Corollary \ref{coro:cn-bound}}
\label{app:proof-of-cn-bound}

For any $(U,V)$ satisfying Eqs.~(\ref{public-lower}) 
and (\ref{markov-condition}), we have
\begin{eqnarray*}
I(V;Y) + I(U;Y|V) 
 &=& I(U,V; Y) \\
 &=& I(U;Y) + I(V;Y|U) \\
 &=& I(U;Y),
\end{eqnarray*}
where the last equality follows from the
fact that $V$, $U$, and $Y$ form a Markov chain.
Similarly, we have
\begin{eqnarray*}
I(V;Z) + I(U;Z|V)
 = I(U;Z).
\end{eqnarray*}
Since we have
\begin{eqnarray*}
I(V;Y) \ge I(V;Z)
\end{eqnarray*}
for degraded sources, we have
\begin{eqnarray*}
I(U;Y|V) - I(U;Z|V) \le I(U;Y) - I(U;Z),
\end{eqnarray*}
which implies the assertion of the corollary. \qed

\section{Miscellaneous Facts}

For reader's convenience, we review some basic
facts on jointly Gaussian random variables.

\subsection{Derivations of Eqs.(\ref{eq:gauss-decompose-1})--(\ref{eq:gauss-decompose-3})}
\label{derivation-of-gaussian-decompose}

The probability density function of
zero-mean Gaussian random vector $\mathbf{X}$ is 
uniquely determined by its covariance matrix $\Sigma_{\mathbf{X}}$.
Furthermore, for any non-degenerate matrix $A$, the covariance
matrix of the Gaussian random vector 
$\mathbf{X}^\prime = A \mathbf{X}$ is given by
$\Sigma_{\mathbf{X}^\prime} = A \Sigma_{\mathbf{X}} A^T$.
On the other hand, any non-degenerate symmetric matrix of the form
\begin{eqnarray*}
M = \left[ \begin{array}{cc}
A & B \\
B^T & C
\end{array} \right]
\end{eqnarray*}
can be decomposed as
\begin{eqnarray*}
M &=& \left[ \begin{array}{cc}
I & 0 \\
B^T A^{-1} & I
\end{array} \right] \\
&& ~~~\left[ \begin{array}{cc}
A & 0 \\
0 & C - B^T A^{-1} B 
\end{array} \right]
\left[ \begin{array}{cc}
I & A^{-1} B \\
0 & I 
\end{array} \right].
\end{eqnarray*}
By using these facts, we can derive 
Eqs.(\ref{eq:gauss-decompose-1})--(\ref{eq:gauss-decompose-3}).

\subsection{Proof of Lemma \ref{lemma:degradable}}
\label{proof-of-lemma:degradable}

By using facts in Appendix \ref{derivation-of-gaussian-decompose},
we can write
\begin{eqnarray*}
Y &=& \Sigma_{xy} \Sigma_x^{-1} X + N_y, \\
Z &=& \Sigma_{xz} \Sigma_x^{-1} X + N_z,
\end{eqnarray*}
where $N_y$ and $N_z$ are Gaussian random variables that
are independent of $X$ and
the variances are
$\Sigma_y - \Sigma_{xy}^2 \Sigma_x^{-1}$ and
$\Sigma_z - \Sigma_{xz}^2 \Sigma_x^{-1}$ respectively.
When Eq.~(\ref{case-1}) is satisfied,
by setting $\bar{X} = X$, $\bar{Y} = Y$, and
\begin{eqnarray*}
\bar{Z} = \Sigma_{xz} \Sigma_{xy}^{-1} \bar{Y} + \hat{N}
\end{eqnarray*}
for Gaussian random variable $\hat{N}$ with 
variance $\Sigma_z - \Sigma_{xz}^2 \Sigma_y \Sigma_{xy}^{-2}$,
we obtain jointly Gaussian sources satisfying
the assertion of the lemma, 
\textchange{where $\hat{N}$ is 
independent of the other random variables.}
When Eq.~(\ref{case-2}) is satisfied with strict 
inequality, we can obtain jointly Gaussian sources
satisfying the assertion of the lemma in a similar manner.
When Eq.~(\ref{case-2}) is satisfied with equality,
by setting $\bar{X} = X$, $\bar{Z} = Z$, and
$\bar{Y} = \Sigma_{xy} \Sigma_{xz}^{-1} \bar{Z}$, 
we obtain sources (not jointly Gaussian because the
covariance matrix is degenerated)
satisfying the assertion of the lemma. \qed


\section*{Acknowledgment}

The first author would like to thank 
Prof.~Ryutaroh Matsumoto
for valuable discussions and comments.
We also thank the anonymous reviewers for their
constructive comments and suggestions.
This research is partly supported by 
Grant-in-Aid for Young Scientists (Start-up):
KAKENHI 21860064.


\profile{Shun Watanabe}{received the B.E., 
M.E., and Ph.D.\ degrees from Tokyo Institute of Technology
in 2005, 2007, and 2009 respectively. He is currently
an Assistant Professor in the Department of Information 
Science and Intelligent Systems of  University of Tokushima.
His current research interests are in the areas of
information theory, quantum information theory,
and quantum cryptography.}

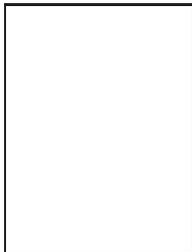
\profile{Yasutada Oohama}{was born in Tokyo, Japan in 1963. He
received the B.Eng., M.Eng., and D.Eng. degrees in mathematical
engineering from University of Tokyo, Tokyo, in 1987, 1989, and 1992,
respectively. From 1992 to 2006, He was with Kyushu University,
Fukuoka, Japan. Since 2006 he has been with University of
Tokushima, Tokushima, Japan. He is currently a professor
at the Department of Information Science and Intelligent Systems.
His current research interest includes basic problems in information
theory and related areas.
}


\end{document}